\documentclass[epj,twocolumn]{webofc}
\usepackage{graphicx}
\graphicspath{ {images/} }
\usepackage[rightcaption]{sidecap}
\usepackage{amsmath}
\usepackage{hyperref}
\usepackage{array}% For tabluar specification
\usepackage{caption}
\usepackage{booktabs}
\usepackage{multirow}
\usepackage{url}

\usepackage[varg]{txfonts}   % Web of Conferences font
\wocname{MPGD,2015}
\woctitle{MPGD,2015}
\begin{document}
\title{Candidate eco-friendly gas mixtures for MPGDs}

 %////////////////////////////////////////////////////////////////////////////////////////////////////////////////////////////////////// Authors //////////////////////////////////////////////////////////////////////////////////////////////////////////////////////////////////// 
    
\author{ L. Benussi \inst{1}
\and S. Bianco \inst{1}
\and G. Saviano \inst{1,2,}
\thanks{Corresponding author; Email address: $Giovanna.Saviano@uniroma1.it$  -and-
$Giovanna.Saviano@cern.ch$}
\and S. Muhammad \inst{1,2,3}
\and D. Piccolo \inst{1}
\and F. Primavera \inst{1}
\and M. Ferrini \inst{2}
\and M. Parvis \inst{4}
\and S. Grassini \inst{4}
\and  S. Colafranceschi \inst{5}
\and J. Kj{\o}lbro \inst{6}
\and A. Sharma \inst{5}
\and D. Yang\inst{6}
\and G. Chen \inst{6}
\ and Y. Ban \inst{6}
\and Q. Li \inst{6}}
     
%///////////////////////////////////////////////////////////////////////////////////////////////////////////////////////////////////////////////// Institute ////////////////////////////////////////////////////////////////////////////////////////////////////////////////////////////////

\institute{Laboratori Nazionali di Frascati - INFN, Frascati, Italy 
\and
University of Rome “La Sapienza” (IT) - Facoltà di Ingegneria, Ingegneria Chimica Materiali ed Ambiente, Italy
\and
 National Center for Physics, Quaid-i-Azam University Campus, Islamabad, Pakistan
\and
Politecnico di Torino-Dipartimento di Fisica (DIFIS) Corso Duca degli Abruzzi, 24, I-10129 Torino, Italy
\and
CERN, Geneva, Switzerland
\and
Peking University, Beijing, China}

%/////////////////////////////////////////////////////////////////////////////////////////////////////////// Abstract //////////////////////////////////////////////////////////////////////////////////////////////////////////////////////////////////////////////////////////////          
\label{sec-2}
\abstract{%
Modern gas detectors for detection of particles require F-based gases for optimal performance. Recent regulations demand the use of environmentally unfriendly F-based gases to be limited or banned. This review studies properties of potential eco-friendly gas candidate replacements.
}

%///////////////////////////////////////////////////////////////////////////////////////////////////////////////////// Introduction ////////////////////////////////////////////////////////////////////////////////////////////////////////////////////////////////////////////////
%
\maketitle
\section{Introduction}
Many currently used refrigerant gases have a great impact on the environment since they either contribute largely to the greenhouse gas effect, or because they tear the ozone layer, or both. In an attempt to protect the environment, regulations preventing the production and use of certain refrigerant gases have been implemented \cite{Benussi:2015gva}. Gas detectors are wide spread for detection, tracking and triggering of charged particles such as muons in Nuclear and High Energy Physics (HEP). A large part of gas muon detectors used in HEP operates with mixtures containing the regulated refrigerants as quenching medium in applications where excellent time resolution and avalanche operation are necessary. Therefore, actions towards finding new mixtures must be undertaken. Gas Electron Multiplier (GEM) \cite{Sauli:1999ya} detectors operate in experiments such as CMS (Compact Muon Solenoid) at the LHC (Large Hadron Collider) with an argon/$CO_2$  mixture \cite{Sharma:2012zzb}. However, for high time resolution applications an argon$/CO_2/CF_4$ mixture is used \cite{Alfonsi:2006ny}, where $CF_4$ has a Global-Warming Potential (GWP) of 7390 \cite{GWP:1}. Resistive Plate Counters (RPC) \cite{Santonico:1981sc} currently operate with a F-based R134a/Isobutane/$SF_6$ gas mixture, with typical GWP of 1430. Investigations into new gas mixtures have to be performed in order to keep the mixture properties while complying with the regulations. A few industrial refrigerant industrial replacements were proposed \cite {GWP:2} as alternatives to R134a. A study of transport properties of currently used gas mixtures in HEP, and evaluation of transport properties of freon-less gas mixtures, was recently published.

The aim of this paper is to discuss some of the important properties of gases for particle gas detectors, to list and summarize basic properties of eco-friendly refrigerants from the literature, to discuss their properties for materials compatibility and safe
use, and make a prediction on selected parameters crucial for the performance of gas detectors considered. While this study is aimed to GEM and RPC detectors, its findings can be considered for selection of ecogas replacement for other gas detectors.

  %///////////////////////////////////////////////////////////////////////////////////////////////////////////////////////////////////////////////////////////////////////////////////////////////////////////////////////////////////////////////////////////////////////////////////////////////
\section{Gas properties}
For a gas mixture to be appropriate in a gas detector, first of all it has to comply with the regulations. Furthermore, its properties must also be appropriate for the specific type of detectors. For example, a gas that is suitable for the RPC detectors may not be
fully optimized for the GEM detectors. To better find the appropriate gas for a detector, an understanding of the influence of different parameters is required. This section aims to clarify the most essential parameters for gases.

In order to estimate the impact of a refrigerant on the environment, the effects have to be quantified. Two important effects are the contribution to the greenhouse effect and the depletion of the ozone layer. The first mentioned effect is measured in Global-
Warming Potential (GWP), and is normalized to the effect of $CO_2$ (GWP = 1), while the effect on the ozone layer is measured in Ozone Depletion Potential (ODP), normalized to the effect of $CCl_3F$ (ODP = 1). The effects of selected refrigerant candidates are
listed in table 1.
When a particle passes through a medium, energy is transferred from the particle to the surroundings.  The energy lost is typically defined as the stopping power expressed as $\frac{1}{ \rho} \left(\frac{dE}{ dx} \right)$ where $\rho$  denotes the density of the medium, E denotes energy, and x is length.The radiation length $X_0$ is a characteristic length of a medium. It describes both the mean distance required for a high energy electron to lose all but $e^{-1}$ of its energy due to bremsstrahlung, and $\frac{7}{9}$ of the mean free path of a $e^+ e^-$  produced by a high-energy photon \cite{RevModPhys.49.421}.
When an incoming particle passes through a medium, it will eventually interact with the medium and transfer some of its energy to ionize atoms. In this process, a pair consisting of an ionized atom and a free electron is produced. The number of ionizations produced by an incoming particle per unit length is denoted by$ N_P$ , in units of $cm^{-1}$ . Each produced ion pair will have an initial kinetic energy and can itself produce an ion pair, called secondary ion pair production. 
In the absence of electric field, electrons move randomly in all direction having an average thermal energy$\frac{3}{2KT}$. In presence of an electric field, the electrons start to drift along the field direction with mean drift velocity $v_d$ (the average distance covered by the drift electron per unit time).
The average distance an electron travels between ionizing collisions is called mean free path and its inverse is the number of ionizing collision per centimetre  (the first Townsend coefficient). This parameter determines the gas gain of the gas.
Many refrigerants may constitute danger for the user and its environment. The greatest dangers involved are the flammability and toxicity. The Health Material Hazardous Material Information System (HMIS), rates Health/ Flammability/ and Physical hazards from 0 (low) to 4 (high). Some refrigerants are incompatible with certain materials, and can either react violently, or have long term effect. Some refrigerants may even produce toxic decomposition and/or polymerization.

The aging phenomenon is very complex and depends on several parameters. The commonly used variables include the cross-sections, electron/photon energies, electrostatic forces, dipole moments, chemical reactivity of atoms and molecules, etc.

%///////////////////////////////////////////////////////////////////////////////////////////////////////////////////////////////////////////////////////////////////////////////////////////////////////////////////////////////////////////////////////////////////////////////////////////////

\section{Estimation of Gas Parameters}
Quantities such as the minimum ionization energy can be found if the stopping power is known. An approximate expression for moderately relativistic particles in the momentum region $0.1\leq\gamma\beta \leq1000$ can be found using the Bethe-Bloch equation, given by

 \begin{equation}
 \frac{1}{ \rho} \left(\frac{-dE}{ dx} \right)= Kz^2 \frac{Z}{A} \frac{1}{ \beta^2} \left(\frac{1}{ 2}  ln \frac{2m_ec^2 \beta^2 \gamma^2 T_{max}}{I^2}-\beta^2- \frac{\delta(\beta \gamma)}{2}\right)
\end{equation}

Where $\left(\frac{-dE}{dx}\right)$  is the mean energy loss per length, $\rho$  denotes the density of the medium, I is the mean excitation energy and $\delta (\beta\gamma)$ is the density effect correction function to ionization energy loss. K is a constant given by $4\pi N_Ar_e^2 m_ec^2r^2$ and $T_{max}$ is the maximum energy transfer in a single collision, given by

\begin{equation}
 T_{max}=\frac{2m_ec^2\beta^2\gamma^2}{1+\frac{2\gamma m_e}{M}+(m_e/M)^2} 
\end{equation}

where M is the mass of the incoming particle. 
The radiation length $X_0$ of an atom can be found by \cite{RevModPhys.49.421}, 
\begin{equation}
X_0 = \frac{716.405(cm^{-2} mol)A}{Z^2(L_{rad}-f(z))+Z\acute{L}_{rad})}
\end{equation}

Table 1 describes the important parameters of some refrigerants. An approximate correlation between primary ionization and atom number has been found based on experimental data by \cite{Smirnov:2005yi}
\begin{equation}
N_p=3.996 \left(\frac{Z_m}{Z^{0.4}}\right)-0.025 \left(\frac{Z_m}{Z^{0.4}}\right) cm^{-1}
\end{equation}

The above expression holds at normal temperature and pressure (NTP) (1atm, $20^0$C). For different pressure and temperature, the number scales with the density. This value should only be taken as a rough estimation though. This formula has proven to work best for hydrocarbons and worst for molecules consisting mainly of fluorine, differing as much as 30\% from the experimental value for $CF_4$. 

\begin{figure}
\centering
\includegraphics[width=1.35\linewidth]{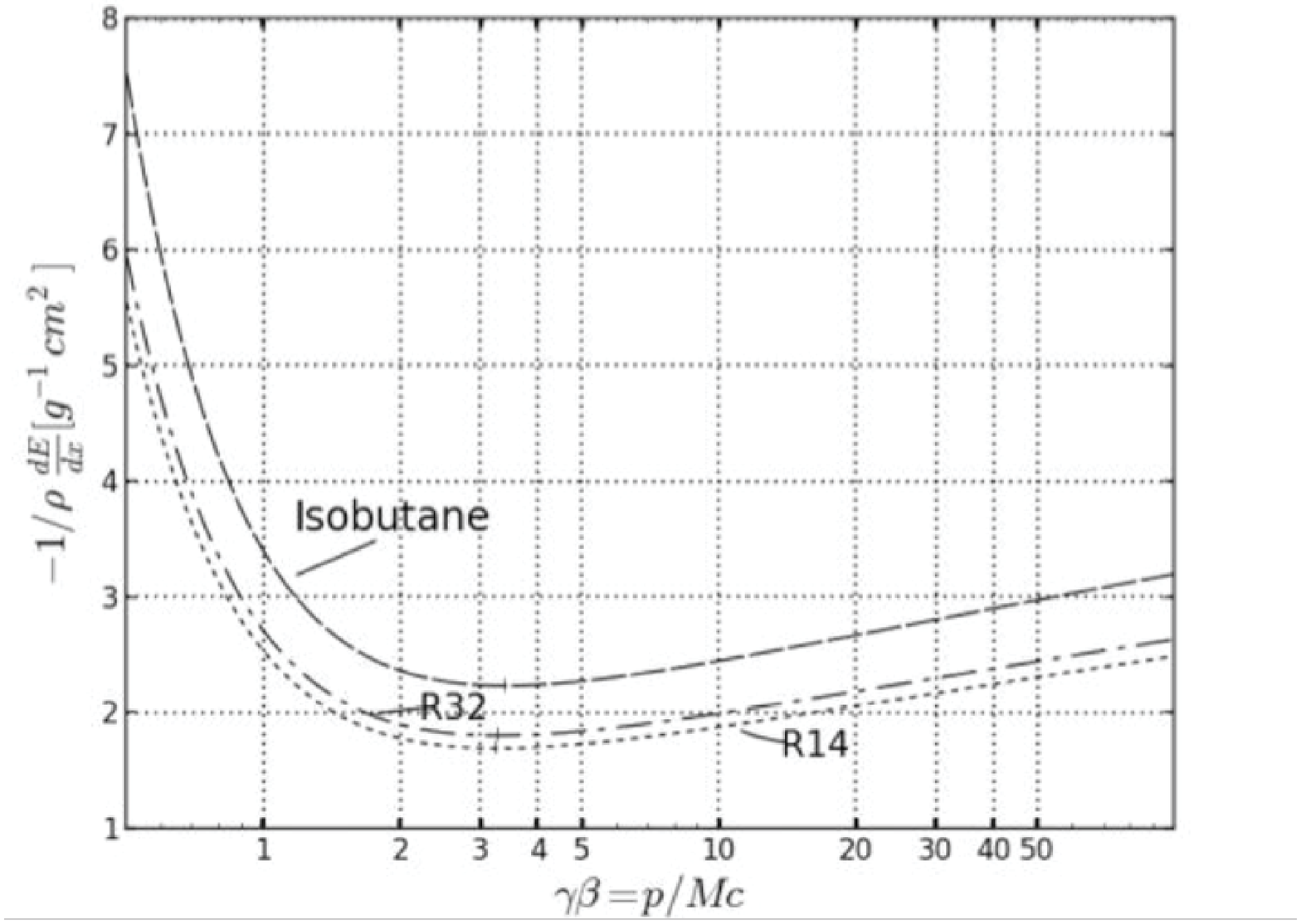}
\caption{Energy loss as a function of the relativistic time dilation factor $\gamma\beta$ for various
refrigerants.}
%\label{fig:Noise}
 \end{figure}

\begin{table}
\caption{Minimum ionization, radiation length and number of primary ion pair creation
for the considered refrigerants, as well as the approximated mean ionization energy
used.}
\begin{tabular}{|l|c|c|c|r|}  \hline
\centering
Name & I & $\left(\frac{-dE}{dx}\right)$ &  $X_0$  &  $N_p$ \\
 & [eV] & $[MeV \frac{g}{cm^2}]$ & $[\frac{g}{cm^2}]$ &$[cm^{-1}]$  \\  \hline
R32 & 89.3602 & 1.80973 & 35.4581 & 49.2  \\ \hline
R7146& 127.401 & 1.67833 & 28.6027 & 92.0 \\ \hline
R600a & 47.848 &2.24057 & 45.2260 & 81.0 \\  \hline
R1234yf & 91.9674 & 1.7734 & 35.8204 & 89.5 \\  \hline
R152a & 78.1889 & 1.88706 & 37.0969 & 67.1 \\   \hline
R1234ze & 91.9674 & 1.7734 & 35.8204 & 89.5 \\ \hline
R115 & 116.695 & 1.69178 & 29.2197 & 98.4  \\  \hline
R1233zd & 106.689 & 1.73915 & 29.7636 & 105 \\  \hline
R290 & 47.0151 & 2.26184 & 45.3725 & 65.2 \\   \hline
R1311 & 271.737 & 1.42486 & 11.5399 & 272 \\   \hline
R134a & 95.0294 & 1.76439 & 35.1542 & 81.6 \\   \hline
R14 & 107.127 & 1.69909 & 33.9905 & 63.6 \\   \hline
R123 & 125.275 & 1.69722 & 25.5416 & 98.4 \\   \hline
R143a & 87.8152 & 1.8126 & 35.8928 & 74.8  \\   \hline
R744 & 88.7429 & 1.81124 & 36.1954 & 37.2  \\  \hline
R23 & 99.9508 & 1.7402 & 34.5214 & 56.9 \\   \hline
R116 & 105.075 & 1.70566 & 34.2947 & 93.3 \\   \hline
RC318 & 101.578 & 1.71721 & 34.8435 & 123 \\  \hline
R218 & 104.13 & 1.70873 & 34.439 & 117 \\  \hline
\end{tabular}
\end{table}

%///////////////////////////////////////////////////////////////////////////////////////////////////////////////////////////////////////////////////////////////////////////////////////////////////////////////////////////////////////////////////////////////////////////////////////////////

\section{Molecules and their optimized geometries}
The freon gases we have been interested in are R134a ($CH_2FCF_3$), R152a ($C_2H_4F_2$), HFO1234ze ($CFHCHCF_3$), HFO1234yf ($CH_2CFCF_3$), $CF_3I$ and HFO1233zd ($CHClCFCF_3$). Meanwhile we choose R12($CCl_2F_2$) to be the standard freon gas model. $CH_4$ and $CF_4$ are the molecules we were used to make a comparison between the $NWC_{HEM}$ calculated results and experimental results, in order to check the reliability of $NWC_{HEM}$. Figure 2 shows the optimized ground state geometries of gas molecules, where the green balls stand for carbon atoms, the grey balls stand for hydrogen atoms, the indigo balls stand for fluorine atoms, the brown balls stand for chlorine atoms, and purple ball for iodine. Figure 3 shows the highest occupied molecular orbitals (HOMO) of gas molecules.
\begin{figure}
\centering
\includegraphics[width=1.25\linewidth]{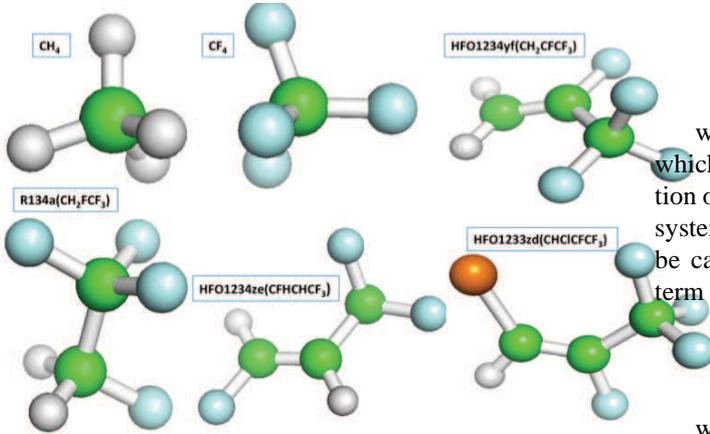}
\caption{Optimized ground state geometries of gas molecules.}
%\label{fig:Noise}
 \end{figure}

\begin{figure}
\centering
\includegraphics[width=1.25\linewidth]{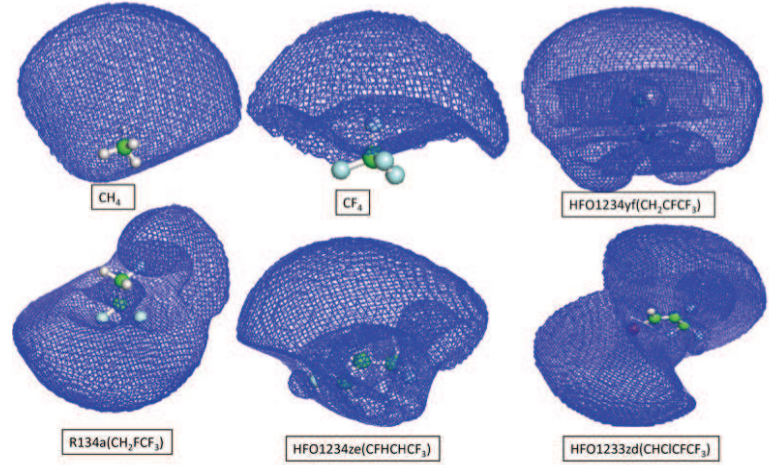}
\caption{Highest Occupied Molecular Orbitals (HOMO) of gas molecules.}
%\label{fig:Noise}
 \end{figure}

%///////////////////////////////////////////////////////////////////////////////////////////////////////////////////////////////////////////////////////////////////////////////////////////////////////////////////////////////////////////////////////////////////////////////////////////////
\section{Calculation of absorption spectrum}
The excitation energy of a molecule is one of the fundamental properties of molecular interactions one can get from experiment. To study these properties, we used the framework introduced in Ref.\cite{K.Lopata:2011} to simulate the time-dependent response of molecules under
external fields using quantum chemical calculations. The framework implemented is the Real-Time Time-Dependent Density Functional Theory (RT-TDDFT) \cite{Y. Takimoto:2007} method within the NWCHEM , making it capable of doing the simulation beyond small perturbation from the ground state.
In our case, we are mostly interested in low excitations of small molecules. We therefore adopt the procedure as described in section 3 of the reference \cite{K.Lopata:2011}. The choice of external field is $\delta$-function-like electric field “kick”

\begin{equation}
 E(t)=k.exp\left [\frac{-(t-t_0)^2}{2\omega^2}\right]\hat{d}
\end{equation}

where $t_0$ is the centre of pulse, $\omega$ is the pulse width, which has dimesnions of time, $\hat{d}=\hat{x}, \hat{y}, \hat{z}$ is the polarization of the pulse, and k is the maximum field strength. The system is then evolved in time, and the dipole moment can be calculated with respect to the added dipole coupling term

\begin{equation}
  V_{\mu\nu}^{app}(t)= - D_{\mu\nu} \cdot E(t)
\end{equation}

where D is the transition dipole tensor of the system. Then we Fourier transform the dipole signals to construct the complex polarizability tensor $\alpha_{i,j}(\omega)$, and finally the dipole absorption spectrum is

\begin{equation}
  S(\omega)= \frac{1}{3} Tr[\sigma(\omega)]= \frac{4\pi\omega}{3c} Tr[Im[\alpha_{i,j}(\omega)]]
\end{equation}

\begin{figure}
\centering
\includegraphics[width=1.4\linewidth]{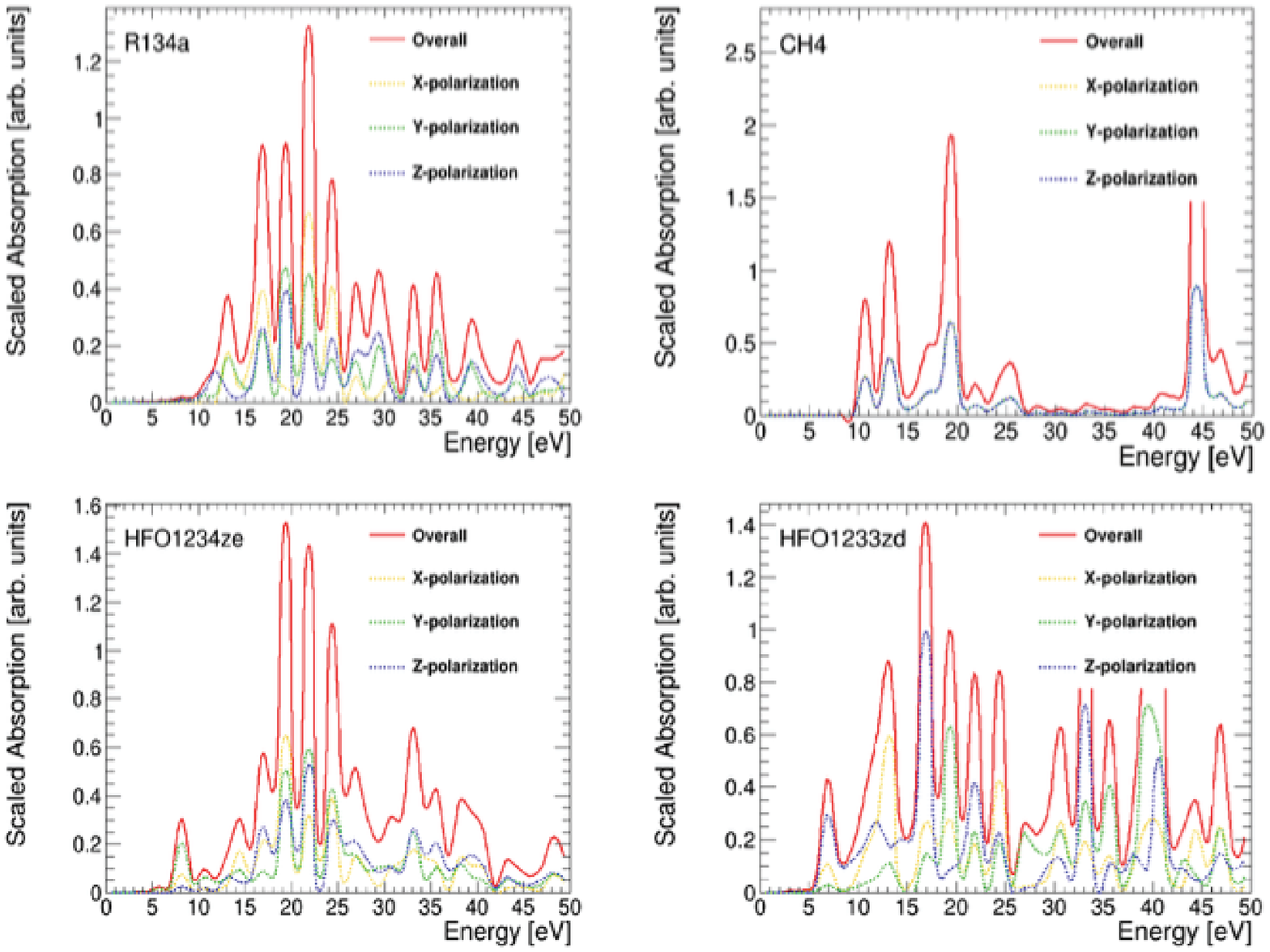}
\caption{Absorption spectrum of interested gas molecules. The calculation is performed in the Atomic Orbital (AO) basis, where the solid line corresponds to the overall dipole strength and the dashed lines correspond to polarizations in the coordinate basis directions.}
%\label{fig:Noise}
 \end{figure}

We first validate our calculation by repeating the calculation of $CH_4$ lowest excitation energy as given in the paper \cite{K.Lopata:2011}. With the same basis set (6-311G) and functional (B3LYP), our calculation gives 11.16 eV excitation energy, which is consistent with the 11.13 eV in the paper. The small difference can come from other minor uncertainty sources like the choice of time separation in the simulation.

%///////////////////////////////////////////////////////////////////////////////////////////////////////////////////////////////////////////////////////////////////////////////////////////////////////////////////////////////////////////////////////////////////////////////////////////////
\section{Estimation of the first Townsend parameter}
Following Ref. \cite{Y. I. Davydov:2006}, we express the dependance of the first Townsend parameter $\alpha$ as a function of the reduced electric field $\frac{E}{p}$, and $\frac{\alpha}{p}= A exp\left[\frac{-Bp}{E}\right]$, The reference \cite{Y. I. Davydov:2006} shows that the first Townsend coefficient at high reduced electric field depends almost entirely on the mean free path of the electrons. The mean free path, which is defined as $\lambda_m$= $\frac{1}{n\sigma}$. where n is the number of atoms per unit volume and $\sigma$ is the total cross section for electron collision with atoms, can be calculated if the environment of gas molecule is provided. For real gases, we cannot take the free path lengths as a constant and the ionization cross section is only a fraction of the total cross section. In that case the estimation needs to be modified and use the following equation:

\begin{equation}
\alpha(r)= Ap\cdot exp\left(\frac{-Bp}{E(r)}\right) \left(1-exp\left(\frac{-I_0n\sigma}{eE(r)}\right)\right)+n\sigma_i\cdot exp\left(\frac{-I_0n\sigma}{eE(r)}\right)
\end{equation}
Numerous measurements of the Townsend coefficients are available for standard gas mixtures, such as those reviewed in \cite{A. Sharma:1992}.

%///////////////////////////////////////////////////////////////////////////////////////////////////////////////////////////////////////////////////////////////////////////////////////////////////////////////////////////////////////////////////////////////////////////////////////////////
\section{Conclusions}
Currently used F-based gases today used in HEP gas detectors are being phased out by industry and replaced by eco-friendly substitute gases. This study has reported on a general survey of industrially available replacements for HEP gases, discussed their
physical properties, materials compatibility and safety issues. Parameters of interest for their use in HEP detectors have been computed following different approaches ranging from parametrizations to quantum chemical calculations: ionization energy, electronegativity, number of primary pairs. Statistical methods to compute amplification parameters of the ionization shower production such as the Townsend coefficients were investigated and preliminary results reported. Promising candidates with lower
GWP are identified for further studies.

%///////////////////////////////////////////////////////////////////////////////////////////////////////////////////////////////////////////////////////////////////////////////////////////////////////////////////////////////////////////////////////////////////////////////////////////////

%\section* {References}


\begin{thebibliography}
\expandafter\ifx\csname url\endcsname\relax
  \def\url#1{\texttt{#1}}\fi

\bibitem{Benussi:2015gva}
L. Benussi et al., “Properties of potential eco-friendly gas replacements for particle detectors”, LNF note INFN-14-13/LNF, CERN-OPEN-2015-004 [arXiv:1505.00701 [physics.ins-det]].

\bibitem{Sauli:1999ya}
F. Sauli and A. Sharma, “Micropattern gaseous detectors,” Ann. Rev. Nucl. Part. Sci. 49 (1999) 341.

\bibitem{Sharma:2012zzb}
A. Sharma, “Muon tracking and triggering with gaseous detectors and some applications,” Nucl. Instrum. Meth. A 666 (2012) 98.

\bibitem{Alfonsi:2006ny}
M. Alfonsi, G. Bencivenni et,al., Aging measurements on triple- GEM detectors operated with CF-4 based gas mixtures,” Nucl. Phys. Proc. Suppl. 150 (2006) 159.

\bibitem{GWP:1}
International Panel on climate changes. http://www.ipcc.ch/publications and data/ar4/wg1/en/ch2s2-10-2.html

\bibitem{Santonico:1981sc}
R. Santonico and R. Cardarelli, “Development of Resistive Plate Counters,” Nucl. Instrum. Meth. 187 (1981) 377.

\bibitem{GWP:2}
M. Abbrescia, Updates from the DESY upgrade meeting, presented at ECFA meeting, June 19th, 2013, DOI http://10.5072/oar.it/1430579044.36.

\bibitem{RevModPhys.49.421}
Y. S. Tsai, “Pair Production and Bremsstrahlung of Charged Leptons,” Rev. Mod. Phys. 46 (1974) 815 [Erratum-ibid. 49 (1977) 521].

\bibitem{Smirnov:2005yi}
I. B. Smirnov, “Modeling of ionization produced by fast charged particles in gases,” Nucl. Instrum. Meth. A 554 (2005) 474.

\bibitem{K.Lopata:2011}
K. Lopata and N. Govind, Journal of Chemical Theory and Computation 7(5), 1344 (2011), URL http://dx.doi.org/10. 1021/ct200137z, http://dx.doi.org/10.1021/ct200137z.

\bibitem{Y. Takimoto:2007}
Y. Takimoto, F. D. Vila, and J. J. Rehr, The Journal of Chemical Physics 127(15), 154114 (2007), URL http://scitation. aip.org/content/aip/journal/jcp/127/15/10.1063/1.2790014.

\bibitem{Y. I. Davydov:2006}
Y. I. Davydov, IEEE Transactions on Nuclear Science 53, 2931 (2006), physics/0409156.

\bibitem{A. Sharma:1992}
A. Sharma and F. Sauli, “A Measurement of the first Townsend coefficient in argon based mixtures at high fields,” Nucl. Instrum. Meth. A 323 (1992) 280.

\end{thebibliography}
\end{document}